\input psbox.tex
\splitfile{\jobname}
\let\autojoin=\relax
\documentstyle[12pt]{article}
\begin{document}
\baselineskip 8truemm
\makeatletter
\input psbox.tex
\makeatother
\begin{titlepage}
\hfill{U\'{S}L-TH-93-02}
\vspace{5 mm}

\begin{center}
{\Large {\bf {Astrophysical aspects of fermion number violation
 in the supersymmetrical standard model}}}\\
\vspace{7mm}
{\bf R.Ma\'{n}ka\footnote{email: manka@plktus11.bitnet,
 internet: manka.phys.us.edu.pl},I.Bednarek\footnote{ internet:
bednarek.phys.us.edu.pl}}\\
\vspace{3mm}
{\sl Department of Astrophysics and Cosmology,}\\
{\sl University of Silesia, Uniwersytecka 4, 40-007 Katowice,
Poland}\\
\end{center}
\setcounter{equation}{0}
\vspace{8 mm}
\centerline{ABSTRACT}
\vspace{2 mm}
The model of the supersymmetrical ball in the supersymmetrical
standard model with additional global U(1) fermion symmetry is
presented.
We show that the supersymmety breaking scale (R-parity), the global
$U(1)$ fermion
symmetry scale and the electroweak symmetry braeking scale are
strictly connected to each other.
The realistic ball with $ M \sim 10^5 - 10^9 M_{\odot}$
and the radius $ R \sim 10^{12} - 10^{14} cm$ is obtained.
Inside the ball all full symmetries are restored.
The ball is stabilized by superpartners and right neutrinos which
are
massless inside.

\vfill
\leftline{PACS number(s): 97.90.+j and 13.15.-f }
\vspace{2 mm}
\end{titlepage}

\section{\bf  Introduction.}
The idea that the supersymmetry is broken for not too large scale
$ M_{ Z} < M_{ SUSY} < 1 TeV$ is promising from the theoretical,
experimental and astrophysical  point of view. This gives the
natural
way to cancel the radiative correction in the perturbative
approach.
The quadrative divergences are still cancel when the supersymmetry
is
softly broken. Supersymmetry predicts the rich structure of the new
particles which could be observed if the supersymmetry breaking
scale
$ M_{ SUSY}$ is not too high. Half of them are boson fields
which
can condensate breaking for example the global symmetries. Such a
case
could happen if the minimal supersymmetrical standard model (MSSM)
\cite{hpn:1984} is
extended by including the right neutrinos and an additional fermion
field.
The aim of this paper is to construct such a model in which the
lepton
number is broken in the result of the sneutrinos condensation
($ {\tilde{\nu}}_{ e} $). This condensation breaks also
the R-parity. The lepton symmetry breaking  for the $\sim TeV$
scale
has also an interesting astrophysical meaning \cite{mh:1990}. It
can produce the neutrino
ball  \cite{bh:1982} \cite{add:1991} with even high mass ($  \sim
10^6 M_{ \odot}$).
The presented supersymmetrical model provides also the "see-saw"
mechanism of the
neutrino mass generation. However, the 1 TeV scale gives the $
\nu_{ e}$ and
$ \nu_{ \mu}$ masses $ \sim 10^{ -3} eV/c^2$, but the unphysically
high
$ \nu_{ \tau}$ ( $ \sim 10 GeV/c^2$). The problem will be solved if
there is no
sneutrinos condensation $ {\tilde{\nu}}_{ \tau}$, or in other words
if the
$ \nu_{ \tau}$ neutrino remainds the Dirac particle.

\section{\bf The  model with the right sneutrino condensation.}
Let us consider the simplest supersymmetrical version
\cite{hpn:1984} of the standard
model where the scalar Higgs sector  is built from  three fields
$ \{ H, \ \overline{H},\Phi \} $. They are in the $ SU_{ L}(2)$
dublet
representation
        \begin{equation}
H       =        \left \{ \begin{array}{ll}
                \ H^{-}    \\
                       H^0
                \end{array}
                \right \}, \
                \overline{H}  =        \left \{ \begin{array}{ll}
                \ H^0    \\
                       H^{ +}
                \end{array}
                \right \}, \
         \end{equation}
and one singlet representation $\Phi$. The left handed leptons are
in
        \begin{equation}
({\bf 1},{\bf 2},-\frac{ 1}{2 })  \sim  L_{ f}  =    \left \{
\begin{array}{ll}
               \ \nu_e    \\
                  e
                \end{array}
                \right \}_L, \
  \left \{ \begin{array}{ll}
               \ \nu_{\mu}    \\
                  \mu
                \end{array}
                \right \}_L    ,\
  \left \{ \begin{array}{ll}
               \ \nu_{\tau} \\
                  \tau
                \end{array}
                \right \}_L ;
         \end{equation}
representation ( $f$ is the family index ).
In this model, however, our attention will be focused on the neutrinos in this
model.
For simplicity we limit ourselves to the first family.
Neutrinos are present in our model in the form of two
supersymmetric
multiplets $ \{ \nu_L, \psi_{\nu_L} \}$ and $ \{ \nu^{c}_R,
\psi_{\nu^c_R} \}$
belonging to the $ SU_{L}(2) \times U_{B}(1)$ representations
$({\bf  2},-\frac{ 1}{2 })$ with the weak izospin $ T_3=\frac{ 1}{2
}$,
the lepton charge $ Q_F=1$ for $ \nu_L $ and  with only
the lepton charge $ Q_F=-1$ for $ \nu^c_R $. Spinors $
\psi_{\nu_L},
\psi_{\nu^c_R} $ eventually will form the Dirac or Majorana fermion
fields
while the sneutrinos $ \nu_L,\nu^c_R $ should gain high masses. But
as
bosons they could also condensate breaking the lepton number
conservation
law. The minimal supersymmetrical standard model is extended by the
additional isosinglet scalar field $S$ carrying the lepton number $
Q_F=-1$.
These singlets can also arise in superstring models.
The Lagrange function may be written as the sum
 \begin{equation}
{\cal L}={\cal L}_{SM}+{\cal L}_{\it ext}
 \end{equation}
of the ordinary Glashow-Salam-Weinberg model with
        \begin{eqnarray}
{\cal L}_{GSW}=-\frac{1}{4} F^a_{\mu \nu} F^{a \mu \nu}
-\frac{1}{4} B_{\mu \nu} B^{\mu \nu}
+    D_{\mu}H^{+} D^{\mu} H + D_{\mu}{\overline{H}}^{+}
D^{\mu}\overline{H}   \nonumber \\
 + D_{\mu}\nu^{+}_L D^{\mu} \nu_L+\partial_{\mu} {\Phi}^{+}
\partial^{\mu} \Phi -
V(H,\overline{H} ,\Phi)+i\overline{e}_R \gamma^{\mu} D_{\mu} L
 + h \overline{e}_R HL
       \end{eqnarray}
with
\begin{equation}
F^a_{\mu \nu}=\partial_{\mu} W^a_{\nu} - \partial_{\nu} W^a_{\mu}
+ig\varepsilon_{abc} W^a_{\mu} W^c_{\nu}
\end{equation}
\begin{equation}
B_{\mu \nu}= \partial_{\mu} B_{\nu} -\partial_{\nu} B_{\mu}
\end{equation}
\begin{equation}
D_{\mu}=\partial_{\mu} -\frac{i}{2} g W^a_{\mu} \sigma^a
-\frac{i}{2} g'
YB_{\mu}
\end{equation}
The potential $V$  in the supersymmetric phase is
fully defined by the the superpotential
\begin{equation}
{\cal W}={\cal W}_{SM} + {\cal W}_{\it ext}
\end{equation}
where the superpotential
\begin{equation}
{\cal W}_{SM} =h_e \varepsilon_{i,j} {\overline{H} }^i L^j e^c_R +
\sqrt{\lambda} ( \varepsilon_{i,j}{\overline{H} }^i H^j \Phi -
\frac{1}{3} {\Phi}^3 )
 \end{equation}
is responsible  for the leptons interaction inside the standard
model.
$R$ parity is a discrete symmetry and if it is an exact symmetry, all particles
 of the
standard model  are R-parity even while their superpartners are R-parity odd.
The relation between $R$ parity, the total lepton number $L$, baryon number $B$
and spin $S$ is as follows
\begin{equation}
R_{p} = (-1)^{3B + L+ 2S}
\end{equation}
$R$ parity can be broken in two ways either explicitly or spontaneously.
In this paper
the lepton and $R$ symmetry violating term is postulated in the
following form
\begin{equation}
{\cal W}_{\it ext} =h_{\nu}  \varepsilon_{i,j} H^i L^j {\nu}^{c}_R
+
\sqrt{\lambda} S {\nu}^{c}_R \Phi
\end{equation}
The supersymmetry breaking potential may be  written as
\begin{equation}
\delta V =
m^2 {\Phi}^{+} \Phi -\lambda u( \varepsilon_{i,j}{\overline{H} }^i
H^j
+ S {\nu}^{c}_R ) \Phi
-\kappa u ( \Phi^3 + {\Phi^{3}}^{+} )
\end{equation}
The total potential is the sum of two parts
\begin{equation}
V_s=V_0 + \delta V_1
\end{equation}
the first one comes from the superpotential ${\cal W}$
\begin{equation}
V_0 = \sum_i | \frac{\partial {\cal W}}{\partial \Phi^i} |^2
\end{equation}
where $ \Phi^{i }=\{H, \overline{H} ,\nu_{ L},{\nu}^{c}_R, S,\Phi
\}$.
Let us suppose that the spontaneous lepton number
breaking accures in the result of the boson condensation.
This means that $H$, $\overline{H}$, $S$ and $ \nu^c_R$ fields get
nonvanishing expectation
 values
\begin{eqnarray}
<H>=\frac{ 1}{\sqrt{2} }\{ 0,v \} ,\  <\overline{H} >=\frac{
1}{\sqrt{2} } \{ v,0 \}, \nonumber \\
, \ <\Phi>=\frac{1}{\sqrt{2}} x, \ <S>=\frac{1}{\sqrt{2}} y
 ,\  <\nu^c_R >=\frac{1}{\sqrt{2}} y
\end{eqnarray}
and most of bosonic and fermionic fields get masses.
This superpotential determines the fermions mass matrix
        \begin{equation}
{\cal M}_{ij}=\frac{ \partial^2 {\cal W}}{\partial \Phi^i \partial
\Phi^j }
 \end{equation}
The part of this matrix concerning  the  $ \Phi^{i }=\{H, \bar{H}, \nu_{
L},{\nu}^{c}_R, S,\Phi \}$     fields has the
following form
$${\cal M'}_{ij}=
      \pmatrix{
0 & m_{ D} & \frac{h_{\nu}}{\sqrt2}y & 0 & 0 & 0   \cr
m_{ D} & 0 &  0 &  0 & \sqrt{\frac{ \lambda}{2 }}x & \sqrt{\frac{
\lambda}{2 }}y    \cr
\frac{h_{\nu}}{\sqrt2}y & 0 & 0& \sqrt{\frac{ \lambda}{2 }}x &  0&
\sqrt{\frac{ \lambda}{2 }}v \cr
0 & 0 &  \sqrt{\frac{ \lambda}{2 }}x & 0 & 0& \sqrt{\frac{
\lambda}{2 }}v \cr
0 & \sqrt{\frac{ \lambda}{2 }}x & 0& 0&0& \sqrt{\frac{ \lambda}{2
}}y \cr
0 &  \sqrt{\frac{ \lambda}{2 }}y & \sqrt{\frac{ \lambda}{2 }}v &
\sqrt{\frac{ \lambda}{2 }}v& \sqrt{\frac{ \lambda}{2 }}y&
-\sqrt{2\lambda}x\cr}$$
where $ m_{ D}=\frac{ 1}{\sqrt{2} } h_{ \nu} v$.
On the tree level the vacuum is determined by the minimum of the
potential
        \begin{eqnarray}
V(v,x,y)=\frac{ 1}{2 } \lambda (y^2+v^2) x^2+
 \frac{ 1}{4 }\lambda (v^2 + y^2 - x^2 )^2 \nonumber \\
 -\frac{ 1}{2\sqrt{2} } \lambda u (v^2 + y^2) x
+\frac{ 1}{2 } m^2 x^2 - \frac{ 1}{2 } u \kappa x^3
\label{wz_8}
       \end{eqnarray}
In fact it depends only on one parameter $ z=\sqrt{v^2+y^2}$. This
means
that we have the flat direction for the potential $ V(v,x,y)$.
One parameter, for example $ v$ must appear in the
result
of the phase transition induced by the radiative correction ( the
Coleman-Weinberg potential \cite{lhr:1987},\cite{pl:1981} ).
Keeping only the contributions associated with the gauge bosons $W$,
$Z$ and the
quark $t$  the radiative corrections give
 \begin{equation}
 V_{ r}(v) = \sum_{ i=W,Z,t,...}\frac{ n_{ i}}{64 \pi^2 } m^{ 4}_{
i}(v)
 \{ ln \frac{ m^{ 4}_{ i}(v) }{Q^2 }-\frac{ 3}{2 } \}
  \end{equation}
  where for example
 \begin{equation}
 m^2_W =\frac{ 1}{4 }g^2 v^2 ,
  \end{equation}
  \begin{equation}
m^2_Z =\frac{ 1}{4 }(g^2+g'^2) v^2 ,
 \end{equation}
  \begin{equation}
  m^2_t = h^2_q v^2, ...
   \end{equation}
$ n_i$ depends on the number of degrees of freedom and the particles
statistics
 \begin{equation}
 n_W=6, \ n_Z=3, \ n_t =-12, ...
  \end{equation}
$Q$ is the renormalization scale. Let us notice that $ V_r (0)=0$.
At the extremun point $ v_0$ where $ \partial V_r (v) /\partial v
=0$ the
potential $ V_r (v_0)$ should also be equal to zero. In other case
it will
produce the nonvanishing cosmological constant in the broken phase
($ v=v_0$).
This estimates the renormalization scale $Q$.
The total potential
 \begin{equation}
 V(v,x,y)=V_r (v) + V_s (z,x)
  \end{equation}
  with $ z=\sqrt{v^2+y^2}$ will have the extremum point at
   \begin{equation}
   \frac{ \partial V}{\partial v }=\frac{ \partial V_r}{\partial v
}
+\frac{ \partial V_s}{\partial z } \frac{ \partial z}{\partial v
}=0
 \end{equation}
  \begin{equation}
\frac{ \partial V}{\partial y }=\frac{ \partial V_s}{\partial z }
\frac{ \partial z}{\partial y }=0
 \end{equation}
  \begin{equation}
  \frac{ \partial V}{\partial x }=\frac{ \partial V_s}{\partial x
}=0
   \end{equation}
   will be determined by the condition
 \begin{equation}
 \frac{ \partial V_r}{\partial v }=0, \  \frac{ \partial
V_s}{\partial x }=0, \
 \frac{ \partial V_s}{\partial z }=0
  \end{equation}
The first extremum condition gives
 \begin{equation}
 v=v_0
  \end{equation}
  $ v_0$ is determined by the electroweak symmetry breaking scale
$v_0  \sim 250 GeV $.
The nontrivial minimum $ V_s$ with respect to $z$ gives\\
  \begin{equation}
 z^2 =\frac{ 1}{\sqrt{2} }u x
 \label{wz_7}
   \end{equation}
Using (\ref{wz_7}) the $ V(v,x,y)$ (\ref{wz_8}) potential  may be rewritten in
 the
following form (Fig.1)
\begin{equation}
V(x) = \frac{ 1}{2 }ax^2 - \frac{ 1}{6 }b x^3 + \frac{ 1}{24 }c x^4
\end{equation}
with
 \begin{equation}
a=m^2 -\frac{ 1}{4 }\lambda u^2
 \end{equation}
 \begin{equation}
b=3\kappa u
 \end{equation}
 \begin{equation}
c=6\lambda
 \end{equation}
The potential $V(x)$ depends only on parameters which
break supersymmetry. In a result the potential $V(x)$ can be
parametrized by one dimensional
constant $u$. The solution with $x = 0$ corresponds the
supersymmetrical phase,
whereas the solution with $x \ne 0$ the phase with broken
supersymmetry.
The parameter $v_0$ is fixed by the W-mass ($v_0 \sim 250 GeV$).
To have the vanishing cosmological constant in the broken phase
for small $B$ and
$(v_0 \ne 0)$  we should choose
\begin{equation}
m=\sqrt{\frac{\lambda}{4}} u \sqrt{1+2(\frac{\kappa}{\lambda})^2}
\end{equation}
then
    \begin{equation}
    x_0 =\frac{ \kappa u}{\lambda }
\end{equation}
The  field $ \Phi$  gains the mass
\begin{equation}
 m_{\Phi} =\frac{ 1}{\sqrt{2 \lambda} }\kappa u
\end{equation}
For example, for $\kappa=10$ , $ \lambda=0.3$
 and for $v_0=250 GeV$ we have $u=51.5 GeV$,
$x_0 = 1716.7 GeV$, $m=665 GeV$, and $m_{\Phi}=664.7 GeV$.
The equation
\begin{equation}
z=\sqrt{v^2 +y^2}
\end{equation}
joins together the spontaneous electroweak symmetry breaking scale
( $ v \sim v_0 $), the lepton symmetry breaking scale ($y=y_0$) and
the supersymmetry breaking scale ($z \ne 0$). The electroweak
breaking is determined by the  W-mass ($v_0 \sim 250 GeV$). It is
easy to notice that  when the lepton number breaking   takes place,
the scale of the supersymmetry breaking is greater then $v_0$. The
values of this scale will increase with $y$ (the lepton symmetry
breaking scale). As a result, due to the see-saw  mechanism of
the neutrino mass generation  we shall have the lepton symmetry
breaking scale $y$ dependent  the light Majorana neutrino mass (Fig.2).
The heavy Majorana neutrino mass meanwhile is $1 \sim TeV$ order.
It is interesting, that when the lepton symmetry is correct ($y=0$),
we have the massless left Majorana neutrino. In this case, however,
the supersymmetry is broken exactly  on the electroweak symmetry
breaking scale ( $ v \sim 250 GeV $).
The high symmetry phase ($z=0$) implies that all symmetries including
supersymmetry, the electroweak $SU_L(2) \times U_Y(1)$ symmetry
and the global $U(1)$ fermion symmetry  are restored.

\section{\bf The astrophysical meaning.}
Let us consider now the  nonhomogenous $\Phi$ configuration in  the
model.
 It will be described by
 \begin{equation}
{\cal L}=\partial_{\mu}{\Phi}^{+} \partial^{\mu}\Phi -U(\Phi)
 \end{equation}
with
 \begin{equation}
U(\Phi)= \lambda_{
\nu}(\Phi^{+}\Phi)|\Phi-\frac{1}{\sqrt{2}}x_0|^{2}
 \end{equation}
The Lagrange equation gives
\begin{equation}
\Box x=\frac{\partial U}{\partial x}
\end{equation}
In the spherical coordinates this equation takes the form
\begin{equation}
\frac{d^2 x}{dr^2} +\frac{2}{r} \frac{dx}{dr}=
\frac{\partial U}{\partial x}
\label{wz_2}
\end{equation}
As the potential takes
the degenerate form
\begin{equation}
U=\lambda{'} x^2 ( x -x_0)^2
\end{equation}
where $x_{0}$ is obtained from the potential $V$
In the thin wall approximation we neglect the second term in equation
 (\ref{wz_2}).
As a result we obtain one dimensional equation which is easy to
solve.
\begin{equation}
\frac{1}{2} (\frac{dx}{dr})^2 = U
\end{equation}
In this approximation the solution may be described as the ball
with the radius $R$
  \begin{equation}
x       =        \left \{ \begin{array}{ll}
\ \ \ \ \  \ \ \ \ \ \ \ \ 0\ \  \ \ \ \ \ \ \ \ \ \ \ \ \ \ \ \ \
\  \ r \le R \\
 x_0 e^{m_{\Phi}(r-R)} /(1+e^{m_{\Phi}(r-R)}) \ \ r>R
                \end{array}
                \right \}
                \label{wz_3}
         \end{equation}
where
\begin{equation}
m^2_{\Phi} =\frac{ \partial U_{ef}}{dx } |_{x_0}
\end{equation}
defines the scalar field mass. Its inverse $ l =1/m$ is the
coherent
length and measures the ball wall size. For $ m_{\Phi} = 667.8 GeV $
this is
$ l=2.9\ 10^{-17}$ cm. So, the wall is really thin and this
approximation
seems to be reasonable. The scalar ball may be thought of as a constant
solution
inside the ball and the soliton solution representing the wall.
{}From the solution (\ref{wz_3}) one can conclude that inside the ball for $r <
 R$
$x = 0$. That means that inside the ball exists a
high symmetric phase which is equivalent
to the condition $x = 0, y = 0, v = 0$ in which all symmetries are restored:
supersymmetry, the $SU(2) \times U(1)$ symmetry of the standard model
and the lepton symmetry. All particels inside the ball are massless.
Outside the soliton $(x \neq 0)$ we have the low symmetry phase with the broken
supersymmetry, electroweak symmetry and the lepton symmetry.
In this region all fermions get masses which are particulary large for
supersymmetric partners $(\sim 1 TeV)$.
Because inside the ball all fermions are massles whereas inside  they get
large masses, they have the natural tendency to fill the ball.
They will give the stabilizing (repulsive) term in the expression for the
total energy of the whole system which will protect from the gravitational
collapse.
The boson part of the ball energy
 \begin{equation}
{\cal E}_b =\frac{ 4\pi}{3 }B+4\pi \int^{\infty}_R drr^2
\{ \frac{ 1}{2 }(\frac{ dx}{dr })^2 +U_{ef}(x) \}
 \end{equation}
In the thin wall approximation we have
  \begin{equation}
           {\cal E}_b= s R^2+\frac{ 4\pi}{3 }B R^3
\end{equation}
where
\begin{equation}
s=1.045 \sqrt{\frac{ c}{3 }}x^3_0 I
\end{equation}
is the surface tension.
The fermion energy corresponds to the repulsive force comming from
the
Pauli principle
\begin{equation}
{\cal E}_f =\frac{AN^{\frac{ 4}{3}}}{R}
\end{equation}
with
\begin{equation}
A = \frac{4}{3}{\pi}^{2}(\frac{9\pi}{2})^{\frac{4}{3}}{\gamma}^{-\frac{1}{3}}
\end{equation}
where $\gamma$ is a number of degrees of freedom.
The total energy of the ball is equal
\begin{equation}
{\cal E}=\frac{N^{\frac{ 4}{3 }}A}{R} +s R^2+\frac{ 4\pi}{3 }B R^3
\end{equation}
In this model we consider the case when the bag constant $B = 0$ so
in the
expression for total energy remains only the term containing the
surface
tension $s$.
\begin{equation}
{\cal E}=\frac{N^{\frac{ 4}{3 }}A}{R} +s R^2
\end{equation}
For example, for $ x_0 = 1716 GeV
s= 10^5 (GeV)^3$.
Minimizing, $\frac{\partial {\cal E}}{\partial R} =0$ gives
\begin{equation}
R_0=(\frac{A}{2s})^{\frac{1}{3} }N^{\frac{ 4}{9 }}
\end{equation}
and
\begin{equation}
{\cal E}_0=\frac{3A}{ 2 R_0}N^{\frac{ 4}{3 }}
\end{equation}
Because inside the ball all fermions are massles wherease inside  they get
large masses, they have the natural tendency to fill the ball.
They will give the stabilizing (repulsive) term in the expression for the
total energy of the whole system which will protect from the gravitational
collapse.
This picture will be energetically favourable untill the Fermi level
 $\varepsilon_{F}$
doesn't exeed the value of the fermion masses in the broken symmetry phase.
It means that inside the ball fermions which masses outside the ball
are larger then inside (right neutrinos, supersymmetric partners, etc)
 dominate.
Because inside the ball energy of the supersymmetric ground state equals zero
whereas outside the ball the cosmological constant (energy of the ground state)
also equals zero, the potential $U(x)$ describes a soliton solution with
the bag constant $B = 0$ and only with different from zero surface tension.
We can notice a simmilarity to a quark star. In this case there is a
 deconfinement
phase
inside the soliton with $B \neq 0$ which is decisive for macroscopic properteis
of the star. In our case $B = 0$ and the values of the bag mass and radius
are determined only by the surface tension.
We shall have the critical  radius $R_c$ when $R_g=R_0$. As
$R_g=\frac{2
{\cal E}}{M^2_{Pl}}$, we have  the critical fermion number $N_c$
defined as
\begin{equation}
N_c=\frac{ 1}{A}( \frac{M^2_{Pl}}{3 \sqrt{4 s^2}})^3
\end{equation}
        \begin{eqnarray*}
R_c=\sqrt{\frac{3A}{M_{Pl}}}, \ \ \ \ \ \
{\cal M}_c =\frac{1}{2} M^2_{Pl} R_c
        \end{eqnarray*}
The previous numerical parameters give
$N_c=5.27 \times 10^{69}$, $R_c=2.37 \times 10^{14}$ cm and $ M_c
=8.06 \times 10^9\ M $.
The ball mass will  depend as $N^{\frac{8}{9}}$, while the energy
of the
corresponding $N$ free particles will depend
linear on $N$. This suggests that the ball is stable considering the
decay for free
particles.

\section{\bf Discussion.}
In this paper the model of the supersymmetrical ball was created.
The ball is
a result of  sneutrinos the  boson neutrino  partners
nonhomogeneous condensation. The realistic ball with $ M \sim 10^5
- 10^9 M_{\odot}$
and the radius $ R \sim 10^{12} - 10^{14} cm$ is obtained.
Inside the ball all full symmetries are restored.
The ball is stabilized by superpartners and right neutrinos which
are
massless inside.

\newpage
\begin{center}
{\bf Figure captions}:\\
\end{center}

$$
\pscaption{\boxit{\psboxto(10cm;10cm){fig1.ps}}}
{Fig 1. The potential V(x) as a function of x (in GeV).}
$$

$$
\pscaption{\boxit{\psboxto(10cm;10cm){fig2.ps}}}
{Fig 2. The light neutrino mass (in $10^(-12)$ GeV) as a function  of the
 lepton
symmetry scale (in GeV).}
$$

\autojoin
\end{document}